\documentclass[a4paper,twoside]{article}
\newcommand{\affil}[1]{$^{\rm #1}$}
\newcommand{\HI}{H\,{\sc i}}

\usepackage[authoryear]{natbib}
\bibpunct{(}{)}{;}{a}{}{,}
\usepackage{graphicx}
\date{} 
\newcommand{\kms}{\mbox{km\,s$^{-1}$}}
\title{\large\bf\flushleft  Galaxy Groups: Proceedings from a Swinburne University Workshop}
\author{\parbox{\textwidth}{\flushleft
\vspace{-0.5cm}
{\it Virginia A. Kilborn\affil{A,E}, Kenji Bekki\affil{B}, Sarah Brough\affil{A}, Marianne T. Doyle\affil{C}, Ekaterina A. Evstigneeva\affil{C}, Duncan A. Forbes\affil{A}, B\"aerbel S. Koribalski\affil{D}, Matthew S. Owers\affil{B}, Chris Power\affil{A}}\\
\vspace{0.4cm}
{\small \affil{A}\,Centre for Astrophysics and Supercomputing, Swinburne University, Hawthorn, VIC, 3122}\\
{\small \affil{B}\,School of Physics, University of New South Wales, Sydney, NSW 2052}\\
{\small \affil{C}\,Department of Physics, University of Queensland, Brisbane, QLD 4072}\\
{\small \affil{D}\,Australia Telescope National Facility, CSIRO, 
         P.O. Box 76, Epping, NSW 1710}\\
{\small \affil{E}\,Email: vkilborn@swin.edu.au}\\
}}
\begin{document}
\twocolumn[
\begin{changemargin}{.8cm}{.5cm}
\begin{minipage}{.9\textwidth}
\vspace{-1cm}
\maketitle
%
%

\small{\bf Abstract:} We present the proceedings from a 2-day workshop
held at Swinburne University on the 24th-25th of May 2005. The
workshop participants highlighted current Australian research on both
theoretical and observational aspects of galaxy groups. These
proceedings include short 1-page summaries of a number of the talks
presented at the workshop. The talks presented ranged from reconciling
N-body simulations with observations, to the \HI\ content of galaxies in
groups and the existence of ``dark galaxies''. The formation and
existence of ultra-compact dwarfs in groups, and a new supergroup in
Eridanus were also discussed.




\medskip
\medskip
\end{minipage}
\end{changemargin}
]
\small

\section{Introduction}

The majority of galaxies in the Universe lie in groups (Tully 1987;
Eke et al. 2004a). However, the physical processes operating in groups
are poorly understood. For example, to what extent do gravitational
interactions and the intra-group medium determine the morphology and
star formation properties of galaxies residing in and around
groups? Simulations of groups of galaxies have not been reconciled
with the observations: In general, many more satellite galaxies are
predicted than are seen in galaxy groups (e.g. Moore et al. 1999;
Klypin et al. 1999). 

Environment appears to play a key role in the evolution of
galaxies. The star formation rate of galaxies in and around (out to
2-3 virial radii) clusters is lower than that of the field (Lewis et
al. 2002; Gomez et al. 2003). The galaxy density at these radii are
comparable to that of groups. There are several possible mechanisms
for the star formation to be quenched such as removal of the gas via
ram pressure stripping (Abadi, Moore \& Bower 1999; Vollmer et
al. 2001; Kenney, van Gorkum \& Vollmer 2004), gravitational
interactions (Toomre \& Toomre 1972; Vollmer 2003; Bekki et
al. 2005a,b), and galaxy mergers (Zabludoff \& Mulchaey 1998). The
latter two are more likely to occur in galaxy groups where the
relative velocities of galaxies are lower.  The observed drop in star
formation rate at large cluster radii suggests that ram pressure
stripping cannot be solely responsible, as this only occurs in the
dense intra-cluster medium near the center of clusters. Thus the
pre-processing of galaxies in groups might be one of the most
important factors in their evolution.

Observations of neutral hydrogen (\HI) in galaxy groups can provide
information about the processes that have occurred in the groups. In
particular, evidence of previous galaxy interactions is often present
in the neutral hydrogen where none is obvious in the optical (e.g. M81;
Yun, Ho \& Lo 1994). In the dense cluster environment, ram pressure
stripping of \HI\ from gas-rich spiral galaxies is observed (e.g. Vollmer
et al. 2001; Oosterloo et al. 2005), corresponding with the
observation that spiral galaxies near the center of clusters tend to
be \HI\ deficient (Solanes et al. 2001).  There is little corresponding
data on the \HI\ content of spiral galaxies in groups. The \HI\ content of
galaxies in compact groups tends to be deficient according to
Verdes-Montenegro et al. (2001), while Stevens et al. (2004) find no
evidence of \HI\ deficiencies in this environment. The few studies of
loose groups tend to support moderate \HI\ deficiencies, caused by
gravitational interactions (Kilborn et al. 2005; Omar \& Dwarakanath
2005).

Gravitational interactions may also play a part in the formation of a
new class of galaxies, the ultra-compact dwarfs (UCD).  UCDs were
first discovered in the Fornax cluster by Hilker et al. (1999) and
Drinkwater et al. (2000), and while they are unresolved in
ground-based optical images, their spectra show they are located at the
cluster velocity. UCDs have luminosities between that of globular
clusters and dwarf galaxies \citep{bek03}. \citet{bek03} proposed that
UCDs were formed from galaxy ``threshing'' of dwarf galaxies, where
the outer parts of the galaxy were tidally removed, leaving
just the nucleus.


Our Swinburne University workshop addressed some of these current
issues in galaxy groups, and summaries of nine of the presentations
follow.


\newpage
\section{Formation of intergroup and intragroup stellar and gaseous objects}

{\it K. Bekki\affil{A}}\\
\vspace{0.4cm}

{\small \affil{A}\,School of Physics, University of New South Wales, Sydney, NSW 2052}\\
\bigskip

We investigate the formation of apparently isolated \HI\ gas and
HII regions in group of galaxies based on gas dynamical simulations
of disk galaxies in groups.
Details on the disk galaxy models
and TREESPH code adopted in the present study have
already been described in Bekki et al. (2002) and Bekki et al. (2005a,b),
so we give only a brief review here. We investigate the
dynamical evolution of stellar and gaseous components 
(including globular clusters) in an
interacting pair of late-type disk galaxies with the mass ratio
smaller than 0.1 (with the larger disk similar to the Galaxy)
in groups of galaxies.
Star formation is modeled as the Schmidt law (Schmidt 1959) with
the exponent of 1.5 and the threshold gas density for star formation
(Kennicutt 1998) is also included. It is found that 
(1) massive isolated \HI\ clouds with the masses of $10^9 {\rm M}_{\odot}$
can be formed as a result of tidal stripping during galaxy-galaxy interaction
in groups (Bekki et al. 2005a, b),
(2) star formation can occur in some of these isolated
\HI\ gas clouds if the companion galaxies are more massive and have
a larger amount ($5 \times 10^9 {\rm M}_{\odot}$)
of gas initially before tidal interaction (Bekki et al. 2005a),
and (3) intragroup GCs can be formed by tidal stripping (Yahagi \& Bekki 2005). 

\begin{figure}[h]
\begin{center}
\includegraphics[scale=0.2, angle=0]{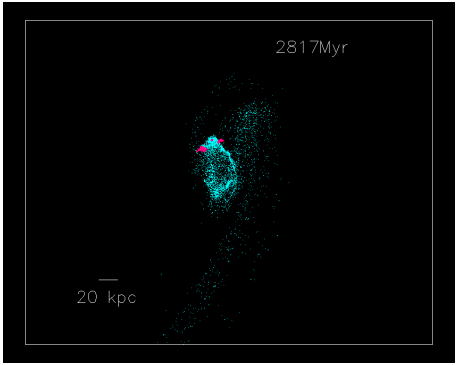}
\caption{The final distributions of gas (cyan) and new stars (magenta)
after tidal stripping of \HI\ gas.}\label{kenji}
\end{center}
\end{figure}

Figure~\ref{kenji} shows the final distributions of gas (cyan) and new
stars (magenta) after tidal stripping of \HI\ gas
from the companion (the smaller galaxy) in the model with the larger disk
mass of $6 \times 10^{10} {\rm M}_{\odot}$. 
The center of the frame coincides with the
center of the larger galaxy and the companion galaxy has gone away from
the larger one so that it can not be seen in this frame. 
It is clear from this figure that new stars can be formed in
the denser part of the intragroup \HI\ ring owing to tidal compression
of the \HI\ gas. Gaseous region around these new stars can be identified
as intragroup HII regions as observed in NGC 1533 by Ryan-Weber et al. (2003).
These results imply that galaxy-galaxy interaction and the resultant
tidal stripping can be responsible  for the formation of
intragroup \HI\ gas clouds and
intragroup HII regions.

\newpage

\vspace{15cm}

\newpage
\section{A Supergroup in 6dFGS}
{\it S. Brough\affil{A}}\\
\vspace{0.4cm}

{\small \affil{A}\,Centre for Astrophysics and Supercomputing, Swinburne University, Hawthorn, VIC 3122}\\
\bigskip


Hierarchical structure formation leads us to presume that clusters of
galaxies are built up from the accretion and merger of smaller
structures like galaxy groups (e.g. \citealt{blumenthal84}).  Although
we observe clusters of galaxies forming along filaments and accreting
galaxy group-like structures (e.g. \citealt{kodama01}) we lack clear
examples of groups merging to form clusters - supergroups.

The Eridanus cloud lies at a distance of $\sim 21$ Mpc and includes
two optically classified groups of galaxies, NGC 1407 and NGC 1332.
These groups are part of the Group Evolution Multiwavelength Study
\citep{osmond04} in which their X-ray properties were analysed.  
The NGC 1407 group shows X-ray emission from intra-group gas,
indicating the presence of a massive structure.  In contrast, the
X-ray emission from the NGC 1332 group is associated with NGC 1332
itself, not with intra-group gas.  \cite{omar05} suggest that there is
intra-group gas associated with NGC 1395, however, no group has
previously been associated with this galaxy.
  

At present there is some debate as to the nature of the Eridanus
cloud, with \cite{willmer89} describing it as a cluster made up of
three-four subclumps and \cite{omar05} describing it as a loose group
at an evolutionary stage intermediate to that of Ursa-Major and
Fornax.  The X-ray information suggests that these are distinct
systems and a possible candidate for a supergroup.



 
To determine what this structure is, it was important to define which
galaxies are associated with which structure.  We obtained galaxy
positions, velocities and $K$-band magnitudes from the 6dFGS DR2
(Jones et al. 2005b) and, as 6dFGS is not yet complete, NED, catalogues for a
circle of radius 15 degrees ($6.6\times6.6$ Mpc), centred on the
position of NGC 1332, in the velocity range 500 -- 2500 km s$^{-1}$,
obtaining 513 galaxies.

\begin{figure}
\begin{center}
\includegraphics[scale=0.3, angle=-90]{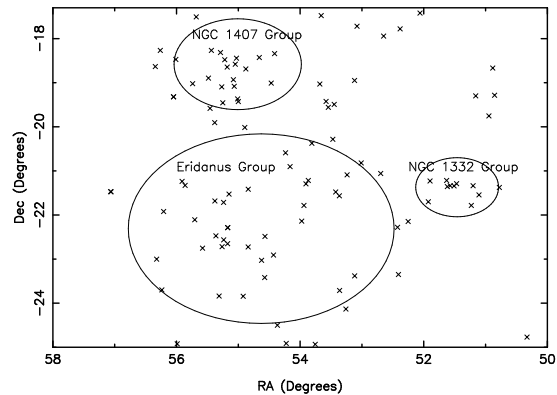}
\caption{
6dFGS \& NED galaxies with $v< 2500$ km s$^{-1}$ and the FOF group
finder output.}
\label{colour_pic}
\end{center}
\end{figure}

We then used the ``friends-of-friends'' (FOF;
\citealt{huchra82}) method to determine which galaxy was associated with 
which structure.  
Figure~\ref{colour_pic} shows that the FOF algorithm finds 3 distinct groups.
The NGC 1407 and NGC 1332 groups are centred on the large ellipticals
of the same name and their X-ray centroids. The Eridanus group is not
centred on any large elliptical or the X-rays associated with NGC
1395.

It is possible to determine the dynamical parameters of these
structures. Consistent with its X-ray information, NGC 1407 is a
massive group of 19 galaxies with a high velocity dispersion,
$\sigma=372$ km s$^{-1}$ and a low crossing time
($t_c=0.03~H_0^{-1}$), consistent with it being virialized.  The NGC
1332 group has fewer galaxies (N = 10) but forms a compact structure
with $\sigma=163$ km s$^{-1}$ and $t_c=0.04~H_0^{-1}$.  The Eridanus
group is made up of more galaxies (N = 31) but is a much looser,
irregular structure.  This is echoed in its low velocity dispersion,
$\sigma=156$ km s$^{-1}$, and high crossing time, $t_c=0.06~H_0^{-1}$.

The NGC 1407 group is a massive group at a late stage in group
evolution, whilst Eridanus and NGC 1332 appear to each be at an early
stage of their evolution, suggesting that this is likely to be a
supergroup.


We also examined the colours of the galaxies with respect to their
density.  The density was measured as the projected surface density of
the 5 nearest neighbours to each galaxy within $\pm 1000$ km s$^{-1}$.
$B$-band magnitudes were obtained from HyperLEDA in order to calculate
$B-K$ colours.  The galaxies are redder in denser environments.  The
density at which this occurs is of the order $\sim 2$ Mpc$^{-2}$,
equivalent to that on the outskirts of this structure.  This indicates
that changes in galaxy properties are occuring at densities equivalent
to those of galaxies currently infalling.

\newpage

\section{\textsc{Hopcat} and Isolated Dark Galaxies}

{\it M.~T.~Doyle\affil{A}, M.~J.~Drinkwater\affil{A}, and D.~J.~Rohde\affil{A}}\\
\vspace{0.4cm}

{\small \affil{A}\,Department of Physics, University of Queensland, Brisbane, QLD 4072}\\

\bigskip

\subsection{Introduction}

The blind HI Parkes All Sky Survey (\textsc{Hipass}), on the Parkes Radio Telescope was completed in 2000.  This survey covers the whole of the southern sky up to Dec. = +2$^{o}$. The \textsc{Hipass} Optical Catalogue (\textsc{Hopcat}) is based on the \textsc{Hipass} catalogue, (\textsc{Hicat}) (Meyer et al.\ 2004 \& Zwaan et al.\ 2004), which represents the largest HI-selected catalogue at this time.  

One motivation for the \textsc{Hipass} survey is to investigate the existance of dark galaxies.  For the purposes of this paper we define a dark galaxy as any HI source that contains gas (and dark matter) but no detectable stars, and is sufficiently far away from other galaxies, groups or clusters such that a tidal origin can be excluded, i.e. isolated dark galaxies.  For further details on \textsc{Hopcat} see Doyle et al. (2005).

\subsection{Optical Galaxies for HICAT}

To identify the optical galaxy match for each \textsc{Hipass} catalogue (\textsc{Hicat}) detection we use 15$\times$15 arcmin SuperCOSMOS images (Hambly et al.\ 2001a; Hambly, Irwin \& MacGillivray 2001b; Hambly et al.2001c) to allow for uncertainity in the original \textsc{Hipass} position. However this creates a problem with matching the correct optical galaxy with its corresponding original \textsc{Hipass} detection due to multiple galaxies present in the majority of the images.  To overcome this problem we cross check \textsc{Hicat} velocity measurements with optical and high resolution radio velocities from NED \footnote{The NASA/IPAC Extragalactic Database (NED) is operated by the Jet Propulsion Laboratory, California Institute of Technology, under contract with the National Aeronautics and Space Administration.} and the 6dF Galaxy  Survey (6dFGS)(Wakamatsu et al. 2003) to validate the galaxy match choice. 

An automated visual interactive program (\textsc{Adric}), where images centred on each \textsc{Hipass} source position are viewed by several people to minimises galaxy selection bias, has been developed. \textsc{Adric} utilizes the SuperCOS images, \textsc{SExtractor}  image analysis (Bertin \& Arnouts 1996), and NED and 6dFGS velocities for cross-checking, to reliably match \textsc{Hicat} HI sources with their optical counterparts. The results for the matching process are shown in Table \ref{Tab:MatchCategoryDescriptions}. \textsc{Hopcat} contains optical galaxy choice categories that not only describe the type of optical galaxy match but the quality of the resulting magnitudes.  

\begin{table}[h]
\begin{center}
\caption{\textsc{Hopcat} optical galaxy matchng results}
\label{Tab:MatchCategoryDescriptions}
\begin{tabular}{lc}
\hline Optical Galaxy Match Category & Percentage \\
\hline 
\itshape{Optically Matched with velocity}		&\\
Single Match &  42\\
Compact group member&16\\
\itshape{Optically Matched with no velocity}	&\\
Single Match &20\\
Compact group member &6\\ 
\itshape{No Matches}	&\\
No Match; Multiple galaxies present&			11  \\
Blank Field; No visible galaxy&	5 	 \\
\hline
\end{tabular}
\end{center}
\end{table}

\subsection{Isolated Dark Galaxies}
The selection criteria used to search for isolated dark galaxy candidates are extinction cut-off and the blank field category. We use an $A_{B_j}$ extinction cut at 1 mag as, beyond this extinction, optically faint galaxies will be dust obscured. 

A total of 3692 galaxies have an $A_{B_j}$ extinction $<$ 1 mag, with only 13 galaxies also in the blank field category. From these, 11 are found to be in over-crowded fields. One object on close inspection does have a faint optical counterpart.  The final dark galaxy candidate has recently been confirmed as a false HI detection.

Our conclusion is that from the 4315 HI radio detections in \textsc{Hicat} no isolated dark galaxies have been found.

\subsection{Summary}
We present the largest catalogue to date of optical counterparts for HI radio-selected galaxies, \textsc{HOPCAT}. Of the 4315 HI radio-detected sources from the HI Parkes All Sky Survey catalogue, we find optical counterparts for 3618 (84 per cent) galaxies.   Isolated `Dark galaxy' candidates are investigated using an extinction cut of $A_{B_j}$ $<$ 1 mag and the 'blank field' optical galaxy match category. We conclude that there are no isolated optically dark galaxies present within the limits of the \textsc{Hipass} survey.

\newpage

\section{Ultra-Compact Dwarf Galaxies in Galaxy Groups}

{\it E.~A.~Evstigneeva\affil{A,B} and Fornax Cluster Spectroscopic Survey team}\\

\vspace{0.4cm}

{\small \affil{A}\,Department of Physics, University of Queensland, QLD 4072}\\

UCDs were first identified by Hilker et al. 1999 (2 objects) and
Drinkwater et al. 2000 (6 objects) during all-object spectroscopic
survey of the Fornax cluster center. The compact objects found in
these surveys, had velocities of cluster galaxies, but were unresolved
in ground-based imaging.  UCDs were subsequently identified in greater
numbers through observations at fainter magnitudes in Fornax cluster
(Drinkwater et al. 2004, Mieske at al. 2004) and Virgo cluster of
galaxies (Jones et al. 2005b).  Possible origins of UCDs include the
following hypotheses: 1) they are luminous intra-cluster globular
clusters; 2) they are extremely luminous star clusters formed from the
amalgamation of stellar super-clusters that were created in galaxy
interactions; 3) they are the remnant nuclei of stripped dwarf
galaxies which have lost their outer parts in the course of tidal
interaction with the galaxy cluster potential -- threshing model; 4)
they are highly compact galaxies formed in the early Universe.

The discovery of UCDs in the Fornax and Virgo clusters raises the
following questions: 1) how common this type of object in clusters in
general? 2) whether the local environment in a cluster affects their
number densities or properties?  3) whether UCDs exist in less dense
environments such as galaxy groups or in the field?  To address these
questions we have carried out a search for similar objects in a range
of galactic environments.

We have chosen six galaxy groups at a redhift similar to that to the
Fornax and Virgo clusters: {\it Dorado, NGC1407, NGC0681, NGC4038,
NGC4697, and NGC5084}.  We tried to select groups with a range of
properties such as number of galaxies and dominant central galaxy
type.  If the threshing model is correct, the efficiency of UCD
formation should scale with the depth of the galaxy cluster
gravitational potential, characterised by the total mass/luminosity of
the system or its dominant central galaxy (Bekki et al. 2003).  Our
sample includes both early and late type dominant central galaxies.
It was interesting for us to explore not only higher density
early-type environments such as e.g. Dorado and NGC1407, but also a
low density spiral environment such as e.g. NGC0681. One of our
selected groups, NGC4038, includes the famous Antennae system -
colliding galaxies, and is very important in understanding UCDs as
supermassive star clusters in interacting systems.

We have carried out a search for UCDs using the 2dF spectrograph on
the Anglo Australian Telescope (AAT) in each group in a single 2
degree field centered on the group center of mass.  To improve our
chances of detecting UCDs in the limited observing time, we defined a
subsample of objects loooking similar to bright Fornax and Virgo UCDs:
they are unresolved in Schmidt photographic plates and have
approximately the same luminosity range as bright Fornax and Virgo
UCDs (-14 $< M_B <$ -11). We also applied a color cut b$_J$ - r $<$
1.7 as no UCDs have been found redder than b$_J$ - r $<$ 1.5.  We had
two observing runs: in November 2004 and in April 2005.  Radial
velocities were determined through cross-correlation with an array of
template spectra. Completeness varied from 24\% to 75\%.

We have found one object (UCD) in Dorado group with the velocity
consistent with it lying in the group.  No UCDs have been found in
five other groups.  We have made rough predictions for how many UCDs
we can find in each group within the chosen luminosity range on the
basis of the Fornax observations.  We adopted two hypotheses: 1) the
number of UCDs scales with the total luminosity of the system; 2) the
number of UCDs scales with the luminosity of the brightest galaxy in
group.  The estimated numbers of UCDs turned out to be different from
the observed ones, suggesting that both of the hypotheses are unlikely
to be correct. However, this conclusion is very sensitive to the group
distance modulus adopted.

The results presented here are very preliminary.  At this point, our
aim was just to see if we can find UCDs in galaxy groups.  To test the
models for the UCD origin, further observations are needed.  We need
to improve the completeness and to observe in a broader magnitude
range to allow for the group distance uncertainties.  Bright UCDs
($M_B<-11.0$) do not seem to exist in large numbers in galaxy clusters
and groups. Searches in a fainter magnitude range will increase the
probability of finding UCDs.  The new AAOmega spectrograph, replacing
2dF on the AAT, will be ideal for these purposes.

\bigskip
The author would like to thank her colleagues on the Fornax Cluster
Spectroscopic Survey who were involved in this project:
M.~J.~Drinkwater, R.~Jurek, P.~Firth, J.~B.~Jones, M.~D.~Gregg and
S.~Phillipps.




\newpage

\section{The GEMS Project and Collapsed Groups}
{\it D. A. Forbes\affil{A}}\\
\vspace{0.4cm}

{\small \affil{A}\,Centre for for Astrophysics \& Supercomputing,
Swinburne University, Hawthorn VIC 3122}

\subsection{GEMS}

The Group Evolution Multiwavelength Study (GEMS) is a survey of
60 nearby groups. Using wide-field
multiwavelength data and mock catalogues, we aim to better
understand the evolution of galaxies in groups and groups
themselves. The groups were selected to have deep ROSAT exposures
at the location of a known (optical) group in the distance range
14 $<$ D $<$ 43 Mpc for H$_o$ = 70 km/s/Mpc (see Osmond \& Ponman
2004 for details). Thus the GEMS groups cover a range of group dynamical
states. \HI\ mapping of GEMS groups is discussed by Kilborn at this
meeting, and Brough presents a dynamical study of two GEMS
groups.

The motivation for this project comes from the fact that groups
are poorly studied relative to clusters and yet contain most
galaxies in the Universe. Recent large surveys like
2dF and SDSS have shown that star formation suppression occurs at
group-like densities.  
But many questions
remain: What is physical mechanism? What is the timescale?


A key property of galaxy groups is the virial radius. This can be
calculated from the velocity dispersion of the galaxies or from
the group Xray temperature (Osmond \& Ponman 2004). 
The combination of \HI\ mapping (e.g. McKay et al. 2004) and 6dFGS spectra
have increased  dramatically the number of confirmed  group members We
are now  in a good position  to compare the virial  radii derived from
velocity dispersion with those  from Xray temperatures. Figure 1 shows
an example for the NGC 5044 group (a massive group). The virial radius
occurs at $\sim$  0.75 Mpc for both methods. Note  the number of group
galaxies beyond 1 virial radius.

\subsection{Isolated Galaxies}

To better understand processes in groups we are also investigating
isolated elliptical galaxies, to act as a control sample. We selected
galaxies with T $<$ --3, D $<$ 130 Mpc, B $<$ 14 and then no
neighbours within 700 km/s, 0.67 Mpc projected radius and 2 B mags
(see Reda et al. 2004 for details). The aim is to understand their
formation, ie are they old `primordial' systems, recent mergers or
collapsed groups?

Our optical imaging study indicated that some isolated galaxies
appear relatively undisturbed while others have much `fine
structure' such as plumes and shells indicating a recent
merger. Most isolated galaxies obey the cluster elliptical
colour-magnitude and fundamental plane relations however again
there were a few notable exceptions. These deviant galaxies
tended to be ones with fine structure and evidence for young
central stellar populations (Reda et al. 2005). A future work
will examine the radial kinematics for isolated ellipticals (Hau
\& Forbes 2005, in prep.). 

\subsection{Collapsed Groups}

The crossing times for some groups are much shorter than the age
of the Universe. So if some groups have collapsed already, how
would they appear? Simulations suggest that multiple mergers will
result in a large, relatively isolated elliptical galaxy with a
group-like Xray halo (the hot gas cooling time is generally
longer than the age of the Universe). Do any of our isolated
galaxies look like collapsed groups?

Our typical isolated galaxy has M$_B$ = --20.5 (and L$_X$/L$_B$ =
30), which is too low luminosity to be an entire group. One
potential collapsed group is NGC 1132 with M$_B$ = --22.0,
L$_X$/L$_B$ = 32 and over 2 B mags brighter than the second
ranked galaxy. It has an old stellar population and is
featureless, suggesting any merger/collapse happened along time
ago.

\begin{figure}[h]
\begin{center}
\includegraphics[scale=0.3, angle=-90]{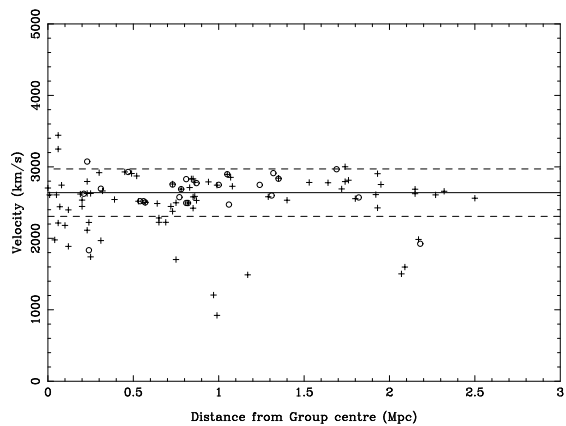}
\caption{Velocity vs distance plot for NGC 5044
group galaxies. The solid
line shows the mean group velocity, and the dashed lines the
velocity dispersion. The virial radius occurs at $\sim$0.75 Mpc. 
}\label{figexample}
\end{center}
\end{figure}

\newpage

\section{Where is the \HI\ in Galaxy Groups?}
{\it V. A. Kilborn\affil{A}}\\
\vspace{0.4cm}

{\small \affil{A}\,Centre for Astrophysics and Supercomputing, Swinburne University, Hawthorn, VIC 3122}\\

\subsection{The GEMS \HI\ Survey}

Deep wide-field neutral hydrogen  observations of galaxy groups
can help us understand the evolution and formation of galaxies within
groups. The existence of \HI\ tails and bridges provide evidence of
previous interactions, and the \HI\ content of galaxies can tell us
about their history. However it is only with the advent of multibeam
receivers in the last ten years that such surveys are now
possible. While HIPASS surveyed the whole southern sky for \HI\
\citep{meyer} its sensitivity for detecting low mass galaxies was
limited to the local volume. Thus to more thoroughly explore the \HI\
content of groups, we have conducted further wide-field \HI\
observations of 16 groups in the Group Evolution Multiwavelength Study
(GEMS; see Forbes, this meeting for a full description). The \HI\
observations were made at the Parkes radiotelescope, using the
multibeam receiver. An area of $\sim 5.5 \times 5.5$ degrees was
observed around each group center. The observations are about twice as
deep as HIPASS, and the velocity resoltution is ten times that of
HIPASS. The sensitivity of the \HI\ survey is $\sim 1-5 \times 10^8
M_{\odot}$, depending on the distance to the group.

\subsection{Results}

We detected 210 \HI\ sources in the 16 datacubes. The positions of the
\HI\ sources were cross correlated with both the NED database, and the
6dFGRS \citep{jones}. We found the majority of the \HI\ detections
corresponded with one or more previously optically catalogued
galaxies, often with known redshifts.  We also found fourteen \HI\
detections that do not match with previously catalogued galaxies. A
further eight \HI\ detections provided the redshift for previously
catalogued galaxies, thus the total number of new group members
accross the 16 groups was 22.  In some groups, there has been an
addition of up to 50\% of known group members, when adding in the new
\HI\ and 6dFGS redshifts.

Optical images of the 14 new \HI\ sources were obtained from the digital
sky survey (DSS).  The majority of the new detections could be
tentatively matched with a visible optical galaxy, however there were
four cases where there was no obvious optical counterpart on the DSS
images. We have obtained high resolution \HI\ images from the ATCA for
one of these \HI\ sources, GEMS\_NGC3783\_12, in the NGC 3783 galaxy
group. The structure of this object is reminiscant of a tidal \HI\
cloud, as it is irregular and has a tail. The formation of this object
is uncertain, and we find no \HI\ bridge to any optical galaxy to the
sensitivity of the datacube ($\sim 5 \times 10^8 M_{\odot}$). The
cloud is $\sim$ 50 kpc in size, and has an \HI\ mass of $\sim 10^9
M_{\odot}$.  There is one further \HI\ cloud candidate in the NGC 3783
group, and we have two more candidates in the NGC 5044 galaxy
group. ATCA observations of these \HI\ cloud candidates are scheduled
for January 2006.

We have begun to investigate the \HI\ content of the group galaxies, and
whether \HI\ deficiencies are seen similar to that in clusters
\citep{solanes} and compact groups \citep{verdes}. We have looked at
one group in depth, NGC 1566. We find the total \HI\ content of the
group is consistent with the optical members of the group. However two
galaxies are about ten times more \HI\ deficient than would be expected
from their optical type and size \citep{kilborn}. There is no diffuse
x-ray emission in this galaxy group, and the \HI\ deficient galaxies do
not lie near the centre of the group, thus we expect the \HI\ deficiency
is caused by tidal stripping of the gas by either the group potential
or other galaxies.

We can place some limits on the existence of intra-group neutral
hydrogen, down to the sensitivity of our survey. Such intra-group \HI\ is
rare, and makes up less than 2\% of the \HI\ detections we found in our
survey. However, our survey cannot place limits on the number of
low-mass \HI\ galaxies in groups.

So to answer the question ``Where is the \HI\ in galaxy groups'' we can
say that to the limit of $\sim 1-5 \times 10^8 M_{\odot}$, it is
mostly contained within galaxies, and \HI\ clouds, tidal tails and \HI\
bridges add only a small part to the total \HI\ content of galaxy groups.

\begin{table}[h]
\begin{center}
\caption{GEMS \HI\ survey results}\label{HItable}

\begin{tabular}{lr}
\hline
No. of groups observed & 16\\
Total galaxies detected & 210 \\
Previously uncatalogued & 14 \\
New Redshifts & 8 \\
Confirmed \HI\ clouds & 1\\
Total new group members & 22\\
\hline
\end{tabular}
\medskip\\
\end{center}
\end{table}


\bigskip
I acknowledge the GEMS team for the general survey, and thank Heath
Jones for supplying the 6dF data.

\newpage

\section{Neutral hydrogen in the M\,83 galaxy group}
{\it B. S. Koribalski\affil{A}}\\

\vspace{0.4cm}

{\small \affil{A}Australia Telescope National Facility, CSIRO, 
         P.O. Box 76, Epping, NSW 1710}\\


ATCA and Parkes \HI\ line observations of the grand design spiral galaxy 
M\,83 (NGC~5236) reveal a very extended \HI\ distribution with a diameter 
of $\sim$100 kpc, several times larger than the optical Holmberg radius. 
While the inner disk of M\,83 rotates remarkably regular, the \HI\ gas 
dynamics appear increasingly peculiar towards the outer regions which show 
clear signs of tidal disruption. The most prominent tidal features are the 
asymmetric outer \HI\ arm bending towards the east of M\,83 and a spectacular 
stellar stream, consisting of mainly old stars, to the north. M\,83 is 
surrounded by numerous dwarf galaxies and, given its large dynamical mass,
is likely to attract and accrete them in regular intervals. \\

\begin{figure}[h]
\begin{center}
 \includegraphics[scale=0.5,angle=0]{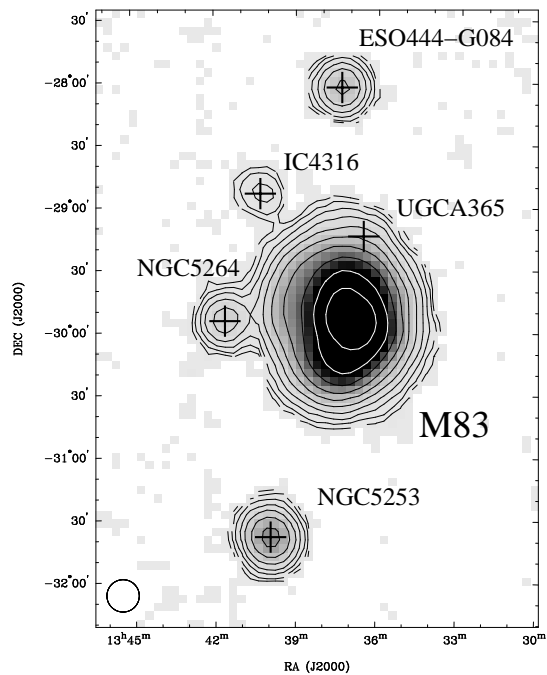}
\caption{\HI\ distribution of the 
    giant spiral galaxy M\,83 (NGC~5236) and its surroundings based on data 
    from the Parkes HIDEEP survey. The contour levels on the 
    left are 0.5, 1, 2, 4, 8, 16, 32, 64, 128, and 256 Jy\,beam$^{-1}$\kms\ 
    (the first contour corresponds to an \HI\ column density of $\sim0.7 
    \times 10^{18}$ atoms\,cm$^{-2}$). We measure a peak \HI\ column density 
    of $5 \times 10^{20}$ atoms\,cm$^{-2}$ occurs at the center of M\,83.
    The gridded Parkes beam of $15.5'$ is indicated at the 
    bottom left. Galaxies detected in \HI\ are marked with a cross.}\label{m83}
\end{center}
\end{figure}

M\,83 (NGC~5236; HIPASS J1337--29) is a late-type spiral galaxy with an 
unusually large \HI\ envelope of $\sim$100 kpc (for $D$ = 4.5 Mpc), at 
least five times larger than its optical Holmberg diameter. It is a member 
of the nearby Centaurus\,A group which appears to consist of two 
subgroupings, one around M\,83 and the other around Cen\,A (NGC~5128).
Figure~\ref{m83} shows the deep Parkes multibeam HI data of the M\,83 subgrouping;
no low-surface brightness \HI\ extensions were detected between M\,83 
and its neighbouring galaxies down to an \HI\ column density limit of 
$\sim10^{18}$ atoms\,cm$^{-2}$ (assuming the \HI\ gas fills the beam).

The \HI\ distribution of M\,83, as revealed with the Australia Telescope 
Compact Array (Koribalski et al. 2005, in prep.) is most remarkable. No longer 
does this grand-design spiral look regular and undisturbed. The \HI\ maps 
show streamers, irregular enhancements, an asymmetric tidal arm, diffuse 
emission, and a thoroughly twisted velocity field, much in contrast 
to its regular appearance in short-exposure optical images. M\,83's \HI\
distribution is enormous, several times larger than its stellar disk. 
It is also a rather massive galaxy, mildly interacting with the neighbouring
dwarf galaxies. The effect of this interaction on the dwarfs can of course
be rather devastating. It is indeed quite likely that M\,83 has accreted
dwarf galaxies in the past. The overall impression of M\,83 in the
high-resolution \HI\ images is that of a distorted one-armed spiral, 
indicating that
it may have interacted or merged with another, smaller galaxy. While the
velocity field in this extended arm appears to follow the general pattern
of rotation, the gas distribution shows numerous irregularities, clumps and
bifurcations.

The eastern-most \HI\ emission of M\,83 which forms part of its peculiar, outer
arm lies $\sim34.5'$ (45 kpc) away from the center of M\,83. We note that 
the dwarf irregular galaxy NGC~5264 lies at a projected
distance of only $25.5'$ (33 kpc) from the eastern \HI\ edge of M\,83. 
Given that the independently measured distances to M\,83 and NGC~5264 are 
very similar, both galaxies are likely to be interacting. 

\newpage

\section{Environmental Analysis of local Starburst Galaxies Selected from the 2dFGRS}
{\it M. S. Owers\affil{A}, C.A. Blake\affil{B}, W.J. Couch\affil{A}, M.B. Pracy\affil{A} and K. Bekki\affil{A}}\\
\vspace{0.4cm}

{\small \affil{A}\,School of Physics, University of New South Wales, Sydney, NSW 2052}\\
{\small \affil{B}\,Department of Physics and Astronomy, University of British Columbia, 6224 Agricultural Road, Vancouver, B.C., V6T 1Z1, Canada}\\

Starburst galaxies are thought to be the progenitors of the rare `E+A'
galaxies (or post-starburst galaxies). \citet{CAB} have used the
2dFGRS \citep{col1,col2} to study the environment of a sample of
nearby `E+A' galaxies. In order to further constrain
triggering/cessation mechanisms for starbursts, we follow the Blake et
al. study and conduct an environmental analysis on a starburst sample
selected from the 2dFGRS.

We use the 2dFGRS spectral line catalogue \citep{lew} to select a
sample of starburst galaxies. We use only high quality spectra from
which we select emission line galaxies with well detected lines.  The
Baldwin, Phillips and Terlevich (BPT; \citet{bpt}) line ratio diagram
was used to differentiate between galaxies where emission is dominated
by star formation from those dominated by AGN activity.  We use the
AGN-starburst demarcation line \citep{kauf} to exclude AGN and define
a star-forming sample.  From this star-forming sample we need to
select a starburst sample, defined as galaxies having normalised star
formation rates (SFR; see \citet{lew}) greater than 10
$\rm{M}_{\odot}$ $\rm{yr}^{-1}$.  We perform an environmental analysis
on this sample including the distance to near-neighbours, projected
surface densities, group membership, correlation with the remaining
2dFGRS galaxies and the luminosity function. The analyses are also
performed on a random sample of 2dFGRS galaxies which act as a
benchmark for comparison to our starburst sample.

The near neighbour (NN) analysis involves finding the transverse
separation of the starburst and its nearest faint and bright
neighbours.  A faint NN is defined as a galaxy with
b*(z)+1\,$<$\,$\rm{b}_{\rm{fn}}$\,$\leq$\,22.5 where b*(z) is the
apparent magnitude an M* ($\rm{M*}=-20.5$, 
\citet{nor}) galaxy would appear as at the redshift of the starburst
and $\rm{b}_{\rm{fn}}$ is the apparent magnitude of the faint NN. A
bright NN is defined as a galaxy with b*(z)+1\,$\leq$\,22.5. The
transverse separation was measured from the starburst to the nearest
faint and bright neighbours, with the NN distance distributions
plotted. The Kolmogorov-Smirnov statistic is used to determine if the
random and starburst NN distributions are drawn from the same parent
distribution. The K-S probabilities measured for both the faint and
bright NN distributions are extremely small, meaning it is highly
unlikely that they are drawn from the same parent distribution as the
random sample.  Since low K-S probabilities indicate a large
difference in two distributions, we can say the faint and bright NN
distributions are quite different from the random sample
distributions. This means a near NN may play a significant role in
triggering a starburst, particularly within 20kpc for a faint NN and
60kpc for a bright NN.

The projected surface density analysis involved measuring the
transverse distance to the fifth nearest bright (as defined above)
neighbour and deriving a surface density from this. The surface
density distribution is compared to the random sample, with the K-S
test showing a vanishingly small probability the random and starburst
samples are drawn from the same parent distribution.  The K-S test
again shows the starburst and random samples differ significantly,
with starbursts showing a difference at higher projected surface
densities meaning tidal interactions with many NN may be significant
in triggering a starburst.

The cross-correlation with the remaining 2dFGRS galaxies involved
measuring the overdensity within comoving spheres of radius 3-15 Mpc
around the starburst galaxy. We compared the starburst sample to a
random sample and an elliptical galaxy sample. The starburst sample
inhabits less dense regions than both the random and elliptical
samples on all scales . This is consistent with the well known
observation that star formation is suppressed in the cores of
clusters.

We cross correlate our starburst sample with the \citet{eke1} 2PIGG
catalogue and determine the corrected total group luminosity (see
\citet{eke2}) for each group containing a starburst. We find that the
starburst galaxies tend to inhabit groups with total luminosities
lower than that of the random sample. That is, starbursts tend to
reside in lower mass groups and avoid rich groups/clusters.

The luminosity function was measured and compared to the whole 2dFGRS
luminosity function. It was found that our starburst galaxies are
preferentially less luminous than the 2dFGRS sample as a whole.

We conclude that: starburst reside in low density environments, the
nearest neighbour may play a role in triggering a starburst whilst
many near neighbours may also induce a burst through tidal
interactions and our starbursts are generally low luminosity objects.


\newpage

\section{The Subhalo Mass Function in Galaxy Clusters : Another Success for the Cold Dark Matter Model?}

{\it C. Power\affil{A} and Brad K. Gibson\affil{A}}\\
\vspace{0.4cm}

{\small \affil{A}\,Centre for Astrophysics and Supercomputing, Swinburne University, Hawthorn, VIC, 3122}\\

\bigskip

One of the defining characteristics of the Cold Dark Matter (CDM) model 
is the hierarchical manner in which massive systems are
assembled through the merging and accretion of less massive progenitors.
However, this merging process is incomplete and the remnant subhaloes 
constitute $\sim 10\%$ of the virial mass in a typical CDM halo,
following a power-law mass function, $N(M) \propto M^{-0.8}$ with $M$ 
the subhalo mass \citep{gao04a}.

In a recent study, \citet{ns04} (NS04) have combined strong and
weak lensing observations of a sample of five rich clusters taken from
\citet{natarajan04} to construct an \emph{averaged} mass function of
the host dark matter halos of the most massive cluster galaxies. They
compared this averaged mass function with the subhalo mass function derived
from the cluster simulations of \citet{springel01}, who studied a single
rich CDM cluster simulated at progressively higher mass and force 
resolution; the best resolved contained $\sim 20$ million particles
within the virial radius. The authors claimed good agreement between
the amplitudes and slopes of the respective mass functions over the
mass range $10^{11} \leq M/{\rm  M_{\odot}} \leq 10^{12.5}$, and
concluded that such concordance provided further observational evidence
in favour of the CDM model.

\begin{figure}
\begin{center}
\includegraphics[scale=0.35, angle=0]{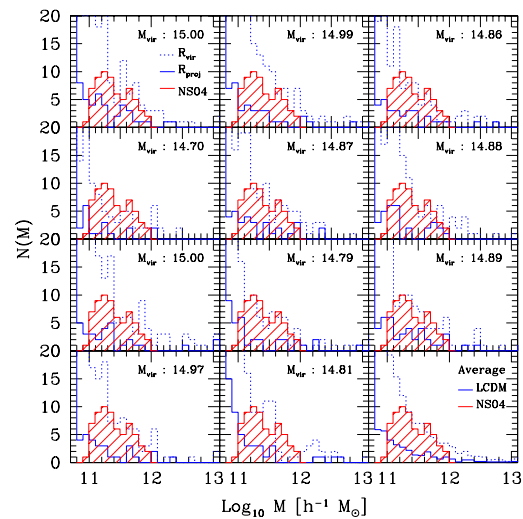}
\caption{\small{Comparison of simulated mass functions with averaged 
      observational mass functions from NS04.}}\label{fig:mf}
\end{center}
\end{figure}

We have investigated the significance of this claim by analysing 
a sample of eleven rich cluster mass CDM haloes, each resolved with 
$\sim 1$ million particles.
In contrast to NS04, we find that subhalo mass functions derived from 
simulated cluster haloes disagree with the mass functions
derived from lensing observations in a systematic manner, as can be seen in 
figure~\ref{fig:mf}. If we consider subhaloes within a projected radius
of $500 h^{-1}$ kpc from the centre of the simulated
cluster\footnote{Comparable to \citet{natarajan04}}, we find that the 
number of a given mass (solid histogram) underestimates the number of cluster
galaxy haloes (hatched histogram) by a factor of $\sim 3$ over the
range of overlap. Agreement between observationally derived and
simulated mass functions is possible only if we consider \emph{all}
subhaloes within the virial radius of the host (dotted histogram).

How significant is this discrepancy for the CDM model? Recent studies 
have demonstrated that galaxies and subhaloes represent different 
populations with distinct spatial and kinematic distributions, and
should not be directly compared \citep{gao04b}. Therefore, it is
unclear that the discrepancy is significant at all without a proper treament
of the galaxy formation process. This is very much work in progress!\\

We warmly thank the Virgo Consortium, Alexander Knebe and Stuart Gill 
for the use of their simulated clusters.





\newpage



\begin{thebibliography}{}

\bibitem[Abadi, Moore \& Bower (1999)]{abadi99}Abadi, M. G., Moore, B., Bower, R. G. 1999, MNRAS, 308, 947 

\bibitem[Baldwin, Phillips, \& Terlevich (1981)]{bpt}Baldwin, J., Phillips, M., Terlevich, R., 1981, PASP, 93, 5

\bibitem[Bertin \& Arnouts (1996)]{1996A&AS..117..393B} Bertin, E., Arnouts, S., 1996, A\&AS, 117, 393 

\bibitem[Bekki et al. (2002)]{bek02}Bekki, K., Forbes, D. A., Beasley, M. A., Couch, W. J. 2002, MNRAS, 335, 1176

\bibitem[Bekki et al. (2003)]{bek03}Bekki, K., Couch, W.J.,  Drinkwater, M.J., Shioya, Y. 2003, MNRAS, 344, 399

\bibitem[Bekki, K. et al. (2005)]{bek}Bekki, K., Koribalski, B. S., Ryder, S. D.,  Couch, W. J. 2005a, MNRAS, 357, L21 

\bibitem[Bekki, Koribalski \& Kilborn (2005)]{bek05}Bekki, K., Koribalski, B. S., Kilborn, V. A. 2005b, MNRAS, in press,
(astro-ph/0505580)

\bibitem[Blake et al.(2004)]{CAB}Blake, C.~A. et al., 2004, MNRAS, 355, 713

\bibitem[Blumenthal et al. (1984)]{blumenthal84}Blumenthal, G.~R., Faber, S.~M., Primack, J.~R., Rees, M.~J., 1984, Nature, 311, 517

\bibitem[Colless et al.(2001)]{col1}Colless, M., et al. 2001, MNRAS, 328, 1039

\bibitem[Colless et al.(2003)]{col2}Colless, M., et al. 2003, astro-ph/0306581

{De~Lucia~G., Kauffmann~G., Springel~V., White~S.~D.~M., Lanzoni~B., Stoehr~F., Tormen~G. \& Yoshida~N. 2004, MNRAS, 348, 333}

\bibitem[Doyle et al. (2005)]{2001MNRAS.361.34-44}Doyle, M.~T., et al., 2005, MNRAS, 361, 34  

\bibitem[Drinkwater et al. (2000)]{dr00}Drinkwater, M.J., Jones, J.B., Gregg, M.D., Phillipps, S. 2000, PASA, 17, 227

\bibitem[Drinkwater et al. (2004)]{dr04}Drinkwater, M.J., Gregg, M.D., Couch, W.J., et al. 2004, PASA, 21, 375

\bibitem[Eke et al.(2004a)]{eke1}Eke, V.~R. et al., 2004a, MNRAS, 348, 866

\bibitem[Eke et al.(2004b)]{eke2}Eke, V.~R. et al., 2004b, MNRAS, 355, 769

\bibitem[Gao et al.(2004a) Gao et al.]{gao04a}{Gao, L., White,~S.~D.~M., Jenkins,~A., Stoehr,~F. \& Springel,~V. 2004a, MNRAS, 355, 819}

\bibitem[Gao et al.(2004b) Gao et al.]{gao04b}{Gao, L., De~Lucia, G., White,~S.~D.~M. \& Jenkins,~A. 2004b, MNRAS, 352, L1}

\bibitem[Gomez et al. (2003)]{gom03}Gomez, P. L. et al. 2003, ApJ, 584, 210

\bibitem[Huchra \& Geller (1982)]{huchra82}Huchra,~J.~P., Geller,~M.~J.,1982, ApJ, 257, 423

\bibitem[Hambly et al. (2001)]{2001MNRAS.326.1315H}Hambly, N.~C., Davenhall, A.~C., Irwin, M.~J., MacGillivray, H.~T., 2001a, 
MNRAS, 326, 1315 

\bibitem[Hambly, Irwin, \& MacGillivray (2001)]{2001MNRAS.326.1295H}Hambly, N.~C., Irwin, M.~J., MacGillivray, H.~T., 2001b, MNRAS, 326, 1295 

\bibitem[Hambly et al. (2001)]{2001MNRAS.326.1279H} Hambly, N.~C., et al., 2001c, MNRAS, 326, 1279 

\bibitem[Hilker et al.  (1999)]{hilker}Hilker, M., Infante, L., Viera, G., et al. 1999, A\&AS, 134, 75

\bibitem[Jones et al. (2004)]{jones}Jones, D. H., Saunders, W., Colless, M. et al.,  2004, MNRAS, 355, 747 

\bibitem[Jones et al. (2005)]{jones05}Jones,~D.~H., Saunders,~W., Read,~M., Colless,~M., 2005a, astro-ph/0505068

\bibitem[Jones et al. (2005)]{jones}Jones, J.B., Drinkwater, M.J., Jurek, R., et al. 2005b, MNRAS, in press

\bibitem[Kauffmann et al.(2003)]{kauf}Kauffmann, G. et al., 2003, MNRAS, 346, 1055

\bibitem[Kenney, van Gorkum \& Vollmer (2004)]{kenney04}Kenney, J. D. P., van Gorkum, J. H., Vollmer, B. 2004, AJ, 127, 3361 

\bibitem[Kennicutt (1998)]{ken98}Kennicutt, R. C., Jr. 1998, ARA\&A, 36, 189 

\bibitem[Kilborn et al. (2005)]{kilborn}Kilborn, V. A., Koribalski, B. S., Forbes, D. A., Barnes, D. G., Musgrave, R. C. 2005, MNRAS, 356, 77

\bibitem[Klypin et al. (1999)]{kly1999}Klypin, A., Kravtsov, A. V., Valenzuela, O., Prada, F. 1999, ApJ, 522, 82

\bibitem[Kodama et al. (2001)]{kodama01}Kodama,~T., Smail,~I., Nakata,~F., Okamura,~S., Bower,~R.~G. 2001, ApJL, 562, 9

\bibitem[Lewis et al.(2002)]{lew}Lewis, I. et al., 2002, MNRAS, 334, 673

\bibitem[McKay et al. (2004)]{mck2004}McKay, N., et al. 2004, MNRAS, 352, 1121

\bibitem[Meyer et al. (2004)]{meyer}Meyer, M. J., et al. 2004, MNRAS, 350, 1195

\bibitem[Mieske et al. (2004)]{mieske}Mieske, S., Hilker, M., Infante, L. 2004, A\&A, 418, 445

\bibitem[Moore et al. (1999)]{moore}Moore, B., Ghigna, S., Governato, F., Lake, G., Quinn, T., Stadel, J., Tozzi, P. 1999, ApJ, 524, 19

\bibitem[Natarajan et al.(2004) Natarajan et al.]{natarajan04}
{Natarajan,~P., Kneib,~J., Smail,~I. \& Ellis,~R. 2004, preprint (astro-ph/0411426)}

\bibitem[Natarajan \& Springel(2004) Natarajan \& Springel]{ns04}{Natarajan,~P. \& Springel,~V. 2004, ApJ, 617, L13 (NS04)}

\bibitem[Norberg et al.(2002)]{nor}Norberg, P. et al., 2002, MNRAS, 336, 907

\bibitem[Omar \& Dwarakanath (2005)]{omar05}Omar,~A. \& Dwarakanath,~K.~S., 2005, JApA, 26, 1

\bibitem[Oosterloo \& van Gorkum (2005)]{oos2005}Oosterloo, T., van Gorkum, J. 2005, A\&A, 437, 190

\bibitem[Osmond \& Ponman (2004)]{osmond04}Osmond,~J.~P.~F., Ponman,~T.~J., 2004, MNRAS, 350, 1511

\bibitem[Reda et al. (2004)]{reda2004}Reda, F., Forbes, D., Beasley, M., O'Sullivan, E., Goudfrooij,
P., 2004, MNRAS, 354, 851

\bibitem[Reda, Forbes \& Hau (2005)]{reda2005}Reda, F., Forbes, D., Hau, G., 2005, MNRAS, in press

\bibitem[Ryan-Weber, Webster \& Bekki (2003)]{ryan03}Ryan-Weber, E., Webster, R.,  Bekki, K. 2003,
in `The IGM/Galaxy Connection: The Distribution of Baryons at z=0', ASSL Conference Proceedings Vol. 281, eds. J. L. Rosenberg and M. E. Putman, Kluwer Academic Publishers, Dordrecht, p. 223

\bibitem[Schmidt (1959)]{sch59}Schmidt, M. 1959, ApJ, 129, 243 \\

\bibitem[Solanes  et al. (2001)]{solanes}Solanes, J. M., Manrique, A., García-Gómez, C., González-Casado, G., Giovanelli, R., Haynes, M. P. 2001, ApJ, 548, 97

\bibitem[Springel et al.(2001) Springel et al]{springel01}
{Springel,~V., White,~S.~D.~M., Tormen,~G. \& Kauffmann,~G. 2001, MNRAS, 328, 726}

\bibitem[Stevens et al. (2004)]{ste04}Stevens, J. B., Webster, R. L., Barnes, D. G., Pisano, D. J., Drinkwater, M. J. 2004, PASA, 21, 318

\bibitem[Toomre \& Toomre (1972)]{toomre1972}Toomre, A. \& Toomre, J. 1972, ApJ, 178, 623

\bibitem[Tully (1987)]{tully1987}Tully, R. B. 1987, ApJ, 321, 280

\bibitem[Verdes-Montenegro et al. (2001)]{verdes}Verdes-Montenegro, L., Yun, M. S., Williams, B. A., Huchtmeier, W. K., Del Olmo, A., Perea, J. 2001, A\&A, 377, 812

\bibitem[Vollmer et al. (2001)]{vol01}Vollmer, B., Cayatte, V., Balkowski, C., Duschl, W. J. 2001, ApJ, 561, 708

\bibitem[Vollmer (2003)]{vol03b}Vollmer, B. 2003, A\&A, 398, 525

\bibitem[Wakamatsu et al.(2003)]{2003ASPC..289...97W} 
Wakamatsu, K., Colless, M., Jarrett, T., Parker, Q., Saunders, W., Watson, F., 2003, ASPC, 289, 97

\bibitem[Willmer et al. (1989)]{willmer89}Willmer,~C.~N.~A., Focardi,~P., da Costa,~L.~N. \& Pellegrini,~P.~S., 1989, AJ, 98, 1531

\bibitem[Yahagi \& Bekki (2005)]{yah05}Yahagi, H, Bekki, K., 2005, submitted to ApJL.\\

\bibitem[Yun, Lo \& Ho (1994)]{yun94}Yun, M. S., Ho, P. T. P, Lo, K. Y. 1994, Nature, 372, 530

\bibitem[Zwaan et al. (2004)]{2004MNRAS.350.1210Z} Zwaan, M.~A., et al., 2004, MNRAS, 350, 1210 


\end{thebibliography}
\end{document}